# Facebook 社交网络图像加密算法

刘晓庆，彭银银，王婕，殷赵霞

（安徽大学 计算机科学与技术学院，合肥 230601）

摘要：Facebook 作为现如今在全世界拥有最大用户量的在线社交网络（online social networks，OSNs）平台，基于 Facebook 社交网络平台的信息保护具有很重要的现实意义。因为用户在社交网络上分享的信息往往以图像为主，本文基于 Facebook 社交网络平台，提出了一种在保证尽可能少的信息损失情况下并且相较以往所提出的加密算法更加安全的图像加密算法。在发送方加密图像上传后，首先其可以抵抗第三方对于加密图像的攻击，防止图像数据泄露，并且在 Facebook 平台对图像进行 JPEG 压缩及滤波等一些未知处理之后，接收方仍然可以解密得到相应的图像数据。

关键字：在线社交网络，JPEG 压缩，图像加密，Facebook

中图分类号：TP309.2　　文献标志码：A

## Image Encryption Algorithm Based on Facebook Social Network

Xiaoqing Liu, Yinyin Peng, Jie Wang, Zhaoxia Yin

(School of Computer Science and Technology, Anhui University, Hefei 230601, China)

**Abstract:** Facebook is the online social networks (OSNs) platform with the largest number of users in the world today, information protection based on Facebook social network platform have important practical significance. Since the information users share on social networks is often based on images, this paper proposes a more secure image encryption algorithm based on Facebook social network platform to ensure the loss of information as much as possible. When the sender encrypts the image for uploading, it can first resist the third party's attack on the encrypted image and prevent the image data from leaking, simultaneously processed by some unknown processing such as compression and filtering of the image on the Facebook platform, the receiver can still decrypt the corresponding image data.

**Key Words:** online social networks, JPEG Compression, Image encryption, Facebook

随着移动互联网的发展，人们社交方式也发生了很大的改变。在现在普遍联系的网络世界中，常用的在线社交网络（OSNs）包括 Twitter、Wechat 以及 Instagram 等。全世界大概有一半的人在使用社交网络，社交网络已经成为日常生活中使用最为广泛的交流与分享方式，并且有越来越多的人习惯在 OSNs 上分享个人的一些日常照片或者相互之间分享一些秘密图像，但是不可避免的是由此带来的隐私保护问题，在大数据分析技术的发展背景下，当个人的日常照片被大量窃取并利用大数据分析，那么泄露的将不仅仅是图像中所暴露的个人隐私信息。在使用社交网络的时候所面临的问题还包括在双方传递隐私图像数据时，如何应对第三方非法获取查看等安全问题。

这些社交网络中 Facebook 又是有着最大用户数量的社交网络平台，所以基于 Facebook 社交网络的隐私保护算法更具有普遍的现实意义。

文章结构为在第 1 部分对社交网络中的隐私保护以及图像加密相关的发展进行简要介绍，在文章的第 2 部分将对所使用的混沌系统进行介绍还有具体的所提出的加密算法，第 3 部分则是将对所提出的加密算法安全性进行分析，最后一部分将对本文做出总结。

## 1 前言

现如今的社交网络中所普遍采用的策略是通过访问控制协议[1-3]，其主要关注在分享的照片只能被特定的用户组访问查看，但也正如[4]中所提出的，简单的分组策略不足以保证信息的进一步安全，而为了保证分组策略的安全性所提出比较复杂的分组保护策略则导致用户难以理解和掌握，体验比较差。本文基于迄今拥有全球最多用户数量的社交网络平台—Facebook，提出了一种图像加密算法，是为了保证图像数据能够安全地传送和接收，并且能够抵抗Facebook对图像的压缩等有损处理。

图像加密算法随着计算机算力的提升，其有了长足的发展，因为图像加密往往随着加密算法复杂性的提升计算代价也会越来越大。基于混沌系统的图像加密算法近年来受到广泛关注，基于混沌系统的图像加密一般是通过随机系统生成伪随机数列，从而控制图像加密中位置置换以及异或加密等基本操作，所以一部分人便关注在提高混沌系统的随机性，从而提出了一些加密算法，如[5,6,7]中其提出的算法往往是将两种或者两种以上的混沌系统耦合或者使用更高维度的混沌系统作为伪随机数生成器，从而生成随机性更强的随机数列。还有基于 DNA 编

码的图像加密算法[8,9,10]，使用压缩技术对图像进行加密[11,12]等等。

为平衡社交网络中用户所需要的安全需求与平台计算之间的关系，本文使用足够满足绝大部分用户的安全性需求的混沌系统作为加密过程所需的伪随机数生成器，并且本文中结合 Facebook 社交网络平台对图像的处理特点提出的图像加密算法，并且在保证图像加密的安全性以及改善由于加密所导致图像信息量的损失条件下而提出一种加密算法。

## 2 图像加密方案

### 2.1 混沌伪随机数生成器

混沌系统作为伪随机数生成器用来生成伪随机数列从而控制图像加密，其主要的优势在于混沌系统对初值就有非常好的敏感性，密钥空间很大。但是为平衡 Facebook 大型社交网络的实际安全需求与计算代价之间的关系，本算法所采用的混沌系统为 Logistic 和 Henon 混沌映射用以分别控制图像加密过程中基本操作—位置置换与异或加密。如图 1 所示 Logistic 映射也叫"虫口模型"，是混沌系统的经典模型，并且对初值有很好的敏感。$x(n+1)=\mu(n)(1-x(n))$，其中 $\mu \in (3.6,4)$，图 2 所示的为 Henon 映射 a=1.4,b=0.3 奇异吸引子图示，Henon 映射是二维迭代系统，其定义如式（1）所示，

$$\begin{cases} x_{n+1} = 1 - ax_n^2 + y_n \\ y_{n+1} = bx_n \end{cases} \quad (1)$$

具有两个控制参数 $a$ 与 $b$，当 $1.05 < a < 1.8, b = 0.3$ 时，Henon 映射处于混沌状态。在[16,17]中其提出的加密算法主要关注在提高混沌系统的随机性，通过二维迭代或更高维度的混沌系统作为伪随机数生成器，但是结合 Facebook 的商用场景与绝大部分用户的安全需求水平，本算法通过两个混沌系统分别控制机密过程的基本操作已能提供足够的安全性要求。用户可以

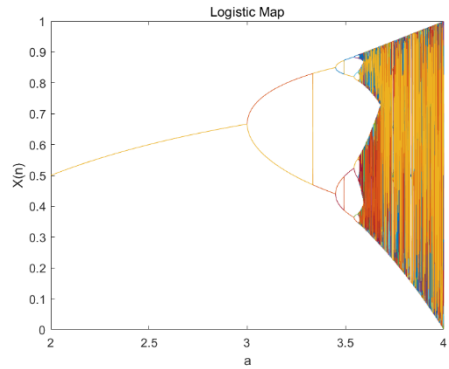

图 1：Logistic 映射仿真图示

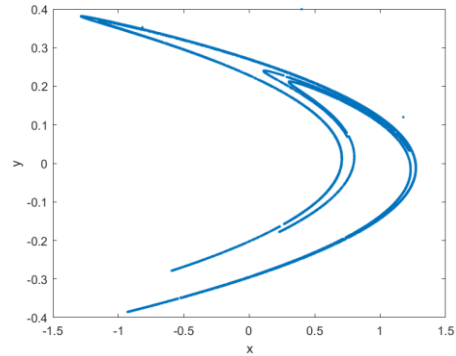

图 2：Honon 映射奇异吸引子图示

通过一对秘钥 $(K_1, K_2)$ 来分作为控制 Logistic 混沌映射与 Henon 映射来生成伪随机数列，利用混沌映射初值极其敏感的特性，既可以为用户提供极大的秘钥空间，也可以极大程度地保证加密的安全性。

## 2.2 图像加密方案

为更好地减少 Facebook 对图像再次压缩导致的图像失真，所以压缩图像的量化表选择为质量因子为 71 的量化表图像进行量化，基本加密流程如图 3 所示，将待加密图像 Img 分割为 $8\times8$ 大小的像素块，再进行

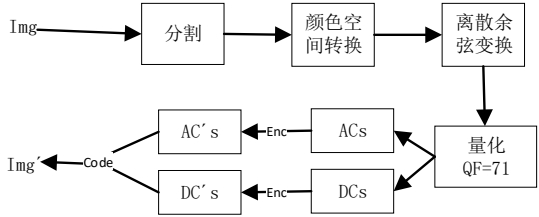

图 3：图像加密基本流程

离散余弦变换以及使用质量因子为 71 的量化表对 DCT 系数进行量化，并且对其 DC 系数和 AC 系数分别采用所提出的加密方案进行加密，通过实验可知 Facebook 对彩色图像压缩采用的采样比例是 4:2:2，故对色度系数矩阵进行加密。最后将加密之后的 DC 系数与 AC 系数进行 Huffman 编码生成密文域 JPEG 图像 Img′。

1) DC 系数加密

DC 系数包含着一张图像的主要能量信息，但是 DC 系数在加密之后往往会导致空域图像的像素值越界问题。在[18]中 He 等人提出了控制量化后 DC 系数范围，从而改善加密后图像像素越界的问题。经过水平移动后的空域图像的像素值 $pv$ 的范围为 $-128 \leq pv \leq 127$，故为尽可能地保证图像在加密之后像素值不超出有效范围，应保证量化后的 DC 系数值 $dcv$ 范围在 $-1024 \leq dcv \leq 1016$。同理由二维离散余弦变换方式可知，DC 系数值在有效范围内并不会保证图像在加密之后不会出现像素越界问题，还与每个图像块中非零 AC 系数值有关。因此为更好地提供加密的安全效果以及改善图像加密之后导致 JPEG 解码后像素值超出有效范围的问题，对 DC 系数提出了以下加密方案：

a) 图像 JPEG 压缩过程中被分成 $8\times8$ 大小的图像块，在 DCT 变换之后每个块的第 1 个位置为 DC 系数，其余 63 个为 AC 系数，故一张大小为 $M\times N$ 的图像会包含有 $\lceil\frac{M}{8}\rceil\times\lceil\frac{N}{8}\rceil$ 个 DC 系数。设一张图像有 n 个 DC 系数依据光栅扫描排列，排列后的集合为 $DCVS=\{DCV_1,...,DCV_i,...,DCV_n\}$，$DCV_i$ 代表依照光栅扫描顺序，第 $i$ 个 DC 系数值。

b) 秘钥 $K_1$ 作为控制 Logistic 映射的 $\mu$ 参数，在 Logistic 映射迭代 1000 次之后进入混沌状后，截取之后的序列作为伪随机数列 $Rl$ 用以控制 DC 系数的加密操作。

c) 首先将同一行内的 $8\times8$ DCT 系数块分为同一组，一张图像有 $\lceil\frac{M}{8}\rceil$ 组，每组内有 $\lceil\frac{N}{8}\rceil$ 个图像块，根据随机秘钥所生成的伪随机数列对每一组内的 DCT 系数块进行位置置换加密。

d) 每次对同一行内的块进行置换加密之后，需要对除哪一行之外的所有块进行置换加密一次。

e) 最后再分别对色度的两个分量按照 c)和 d)进行加密。

2) AC 系数加密

AC 系数存储着图像的一些细节信息，所以对 AC 系数进行进一步的加密。其具体加密过程如下：

a) 由秘钥 $K_2$ 作为 Henon 映射的关键参数 $a$，生成伪随机数列 $R$，并且采用 round(R)函数，将生成的随机数列 R 四舍五入转化成二进制随机数据流 $Rb$。

b) 对二进制随机数据流进行 $Rb$ 进行分组，每 11bit 分为一个加密最小单元（ECU），由 DCT 变换可知 11bit 可以表示所有的非零 AC 系数。将所有的 ECU 按照每 63 个分为一组，记 $ECU_i^j$ 为第 $i$ 组的第 $j$ 个 ECU，$ECU_i^j(k)$ 表示第 $i$ 组的第 $j$ 个 ECU 的前 $k$ bit 二进制数据。

c) 对第 $p$ ($1 \leq p \leq \lceil\frac{M}{8}\rceil \times \lceil\frac{N}{8}\rceil$)块所处第 $q$ 个位置的非零 AC 系数 $cof(p,q)$ 进行进制转换，转化为长度 $l$ 的二进制 $bin(p,q)$，如式(2)所示对其进行加密，加密后为 $bin'(p,q)$，并转换为十进制作为加密后的 AC 系数 $cof'(p,q)$ 进行最后的 JPEG 编码。

$$bin'(p,q) = bin(p,q) \oplus ECU_p^q(l) \quad (2)$$

## 3 图像加密越界分析与处理

在图像上传到 Facebook 社交平台上之后，Facebook 会将图像进行 JPEG 压缩[11]，JPEG 压缩是一种常用的图像压缩方式。JPEG 压缩主要包括图像分块、离散余弦变换、数据量化以及哈夫曼编码过程，其中量化是会导致图像信息损失的过程。在[14]Sun 等人提出的基于 Facebook 加密算法中，通过对 UCID-v2[15]中 1338 张实验图像的结果显示，对于上传到 Facebook 社交网络平台的图像绝大部分被压缩成质量因子为 71 的 jpg 文件。而对于上传的图像大小小于 $2048 \times 2048 pixels$ 时将不对图像再进行处理。

经过图像加密后，在进行 JPEG 解码之后图像的像素值会出现越界问题，由此会产生一些图像的兼容性问题，当图像上传到 Facebook 社交网络平台后，Facebook 会针对图像的空域像素值进行处理进行再压缩，DC 直流系数是 64 个图像像素值的平均值，针对 DC 系数的位置置换加密能够保证直流系数在加密之后，DC 系数不会超出有效范围$(-1024 \leq dcv \leq 1016)$,所以不会出现 DC 系数加密而导致的图像像素值越界的问题。AC 交流系数不会影响图像块整体的平均值，但是会对每点的像素值产生扰动，采用异或加密后使 AC 系数产生改变，仍不可避免地会导致图像的像素值超出有效范围，针对超出有效范围的 AC 系数块，在 Sun 等人[14]的方案中，其对存在像素值超出有效范围的像素块进行越接处理。对于存在越界像素的 DCT 系数块，通过对所有的 AC 系数乘以一个缩小系数 $0 < \alpha < 1$，来将图像的像素值控制在有效范围内。但是这样处理所存在的

问题在于，每一个图像块内的像素相关性很高，故每个图像块的 AC 系数都非常小，而乘以一个缩小系数后，会使很多较小的非零 AC 系数变为 0，从而导致图像的信息损失较大。本算法对于越界像素的像素值，将越界的像素点采取向有效范围内收缩，即对于像素值 $pixel<0$ 的像素点全部空域收缩到 0，而对于像素值 $pixel>255$ 的像素点全部在空域修改为 255。这样可以更好地改善图像的兼容性，避免解码出现错误而导致的文件损坏或者不兼容问题。

## 4　图像解密过程

图像加密之后，通过 Facebook 社交网络平台进行分享传发之后，所有可以查看该图像的用户可以通过密钥对 $(K_1, K_2)$ 对图像进行解密。用户通过 Facebook 社交网络平台获取到秘密图像后，获取图像 Y、Cb 与 Cr 三个色彩分量。在解密过程中需要对每个分量进行分别的解密，但是解密步骤是相同的。首先需要对 AC 系数进行解密，通过秘钥 $K_2$ 控制 Henon 混沌映射生成伪随机数列，并进一步转化成二进制伪随机密钥流，按照 2-2)部分对密钥流分组，并且对所有非零的 AC 系数进行异或解密。当 AC 系数解密之后，再对 DC 系数进行解密操作。

接收方通过秘钥 $K_1$ 控制 Logistic 混沌映射生成随机序列，按照 2-1)部分所提出的加密过程，将 DC 系数恢复到原来的位置。在将三个色彩分量的 DC 系数和 AC 系数解密后，对其进行 Huffman 编码保存查看。

## 5　实验效果与安全分析

在本部分实验是使用 SIPI Image Database 图像数据库的常用测试图像进行展示说明。通过 Facebook 社交网络平台

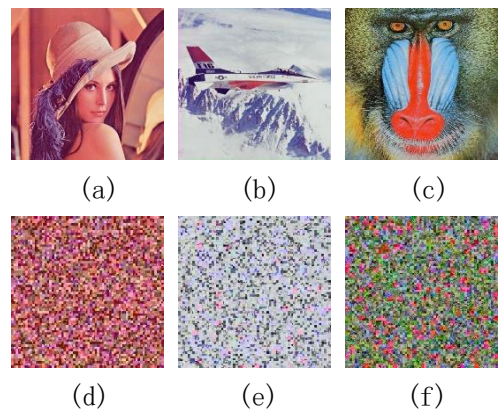

图 4：(a)、(b)与(c)分别是质量因子为 71 的 Lena、Airplane 与 Mandril JPG 图像，(d)、(e)和(f)分别对应(a)、(b)与(c)的加密效果图像。

可以通过对图像空域像素值进行分块、电平移动、DCT 变换以及量化后获取图像的频域 DCT 系数从而进行解密。并且相较于 Sun 等人所提出的[14]加密方案更加安全，可以抵抗在论文[20]中提出的一些针对 JPEG 图像格式的边缘探测攻击方式以及常用的 DCM(DC category mapp-ing)，NCC(nonzero coefficients count)，EAC(energy of AC coeffici-ents)和 PLZ(position of last nonzero coefficients)边缘探测攻击，在表 1 中对 Lena、Airplane 与 Mandril 三张加密后图像的三个色彩分量(Y,Cb 与 Cr)的信息熵，信息熵是信息论中很重要的概念，图像的信息熵最大值为 8，也就意味着当图

像的信息熵无限接近于 8 时，图像所表现的信息特征就更加趋近于随机噪声，同时也就意味加密安全性上更加好一点，所以图像的信息熵也是作为验证加密效果的一个指标。因为人眼对图像的识别主要

表 1：图像分量信息熵

|  | Y | Cb | Cr |
|---|---|---|---|
| Lena | 7.4985 | 5.2962 | 5.2329 |
| Airplane | 6.8006 | 4.4231 | 3.9982 |
| Mandril | 7.1329 | 6.0707 | 6.0486 |

来自于亮度，JPEG 压缩中对于亮度采用全采样方式，故加密之后亮度的信息熵处在 6.8 及以上，说明图像的混乱性与随机噪声很接近。Lena、Airplane 与 Mandril 三张彩色图像在加密之后想的 PSNR 值分别是 15.18、15.71 与 15.24，图像在加密之后其 PSNR 值都在 15 左右，能够在一定程度上说明图像加密的良好效果。

而针对图像加密的安全性讨论，不得不关注加密后图像对抗边缘探测攻击的能力。[14]中所提出的加密算法能够抵抗 DCM、NCC 以及 EAC 等边缘探测攻击，但是其无法抵抗 PLZ 的边缘探测攻击。通过将使用本文所提出的加密算法加密后的彩色图像 Y 分量进行 PLZ 攻击以及[14]所提出的加密方案。图 4 是对使用本算法所加密的 Lena 图像使用 DCM、NCC、EAC 以及 PLZ 边缘探测所示效果，可以看出加密后的图像使用现有的边缘探测攻击无法获取图像的轮廓信息，在图 6 中使用PLZ边缘轮廓探测对使用[14]加密后的 Lena、Airplane 以及 Mandril 进行 PLZ 边缘轮廓探测，由

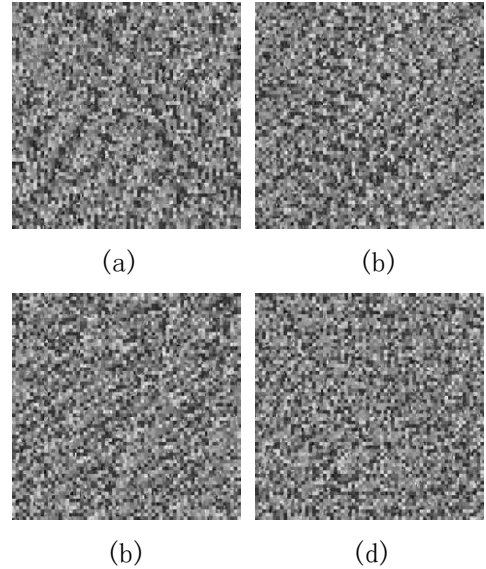

(a) (b)

(b) (d)

图 5：(a)、(b)、(c)与(d)分别是对加密后的 Lena 图像亮度分量使用 DCM、NCC、EAC 与 PLZ 进行边缘探测攻击结果。

于[14]的加密方案并没有改变原始图像最后一位非零 AC 系数的位置，但是在本算法中在 DC 加密过程中，图像块一同执行位置置换加密，打破了非零AC系数的位置特征，加密的安全性得以提高。

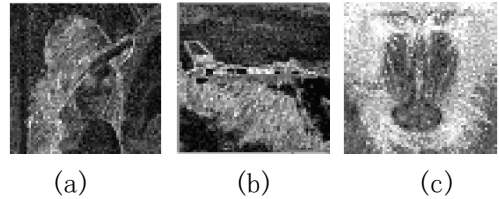

(a) (b) (c)

图 6：(a)、(b)、(c)对使用[14]加密后的 Lena、Airplane 与 Mandril 与应用 PLZ 进行边缘探测攻击结果。

## 6 结论

本文提出了一种基于 Facebook 社交网络平台的图像加密方案，通过对已有的对 Facebook 社交网络平台处理图像的总结，并且随机选取的 BOOS Image Database 中 2000

张灰度图和 UCID-v2[19]上 1338 张彩色图上传到 Facebook 社交网络平台分析其处理方式,并且以此设计了一种基于 Facebook 的彩色图像加密方案,通过实验表明相较于以往的加密方案更好地抵抗边缘轮廓探测,并且采用的混沌映射,用以生成伪随机数列,在不为 Facebook 的大型网络带来巨大运算代价的同时,能够拥有更大的秘钥空间,从而进一步保证图像加密的安全性。

## 7 参考文献